\documentclass{article}

\usepackage{PRIMEarxiv-eusprig}

\usepackage[utf8]{inputenc} 
\usepackage[T1]{fontenc}    
\usepackage[colorlinks,citecolor=blue]{hyperref}       
\usepackage{url}            
\usepackage{booktabs}       
\usepackage{amsfonts}       
\usepackage{nicefrac}       
\usepackage{microtype}      
\usepackage{lipsum}
\usepackage{fancyhdr}       
\usepackage{graphicx}       
\graphicspath{{media/}}     

\usepackage[textsize=tiny, textwidth=60,backgroundcolor=yellow!40, linecolor=yellow, tickmarkheight=10]{todonotes}
\usepackage{comment} 
\usepackage{fancyvrb}

\usepackage[authoryear]{natbib}

\title{When Copilot Becomes Autopilot: \\Generative AI's Critical Risk to Knowledge Work and a Critical Solution
}

\author{
Advait Sarkar \\
Microsoft Research, University of Cambridge, and University College London \\
\texttt{advait@microsoft.com}
\And
  Xiaotong (Tone) Xu \\
  University of California, San Diego \\
  \texttt{xt@ucsd.edu} \\
   \And
  Neil Toronto, Ian Drosos, Christian Poelitz \\
  Microsoft Research \\
  \texttt{\{neil.toronto, t-iandrosos, cpoelitz\}@microsoft.com} \\
}

\begin{document}
\maketitle

\begin{abstract}
Generative AI, with its tendency to ``hallucinate'' incorrect results, may pose a risk to knowledge work by introducing errors. On the other hand, it may also provide unprecedented opportunities for users, particularly non-experts, to learn and apply advanced software features and greatly increase the scope and complexity of tasks they can successfully achieve. 

As an example of a complex knowledge workflow that is subject to risks and opportunities from generative AI, we consider the spreadsheet. AI hallucinations are an important challenge, but they are not the greatest risk posed by generative AI to spreadsheet workflows. Rather, as more work can be safely delegated to AI, the risk is that human critical thinking -- the ability to holistically and rigorously evaluate a problem and its solutions -- is degraded in the process. The solution is to design the interfaces of generative AI systems deliberately to foster and encourage critical thinking in knowledge work, building primarily on a long history of research on critical thinking tools for education.

We discuss a prototype system for the activity of critical shortlisting in spreadsheets. The system uses generative AI to suggest shortlisting criteria and applies these criteria to sort rows in a spreadsheet. It also generates ``provocations'': short text snippets that critique the AI-generated criteria, highlighting risks, shortcomings, and alternatives. Our prototype opens up a rich and completely unexplored design space of critical thinking tools for modern AI-assisted knowledge work. We outline a research agenda for AI as a critic or provocateur, including questions about where and when provocations should appear, their form and content, and potential design trade-offs.
\end{abstract}


\section{Introduction and Background}
Generative AI, defined as any \emph{``end-user tool [...] whose technical implementation includes a generative model based on deep learning''} \citep{sarkar2023eupgenai}, is becoming widely integrated into knowledge work. A prominent contemporary metaphor to describe these new tools is the ``AI Co-Pilot'' \citep{sellen2023rise}, which captures the aspiration that, like human copilots, AI copilots offer support, expertise, and backup, while the (human) pilot remains ultimately in control to make critical decisions.


The discourse around the risks of generative AI revolves principally around \emph{hallucinations}, a nebulous term which we can loosely take to mean generated output (text, code, etc.) that is ``false'' or ``incorrect'' according to some independent objective criterion. Undetected hallucinations can cause serious problems for knowledge work, as in the now-infamous story of the fictitious case citations generated by ChatGPT and submitted by unsuspecting lawyers as part of a legal brief \citep{sarkar2024large}.

However, there is reason to believe that hallucinations, in the form of verifiable factual errors, are not an insurmountable challenge. Several benchmarks show a steady improvement in large language model (LLM) performance with improvements in pre-training datasets and techniques, learning from human feedback, and model size \citep{chang2023survey}. Benchmarks targeted at measuring hallucinations similarly show a reduction in hallucinations \citep{li2023halueval, zhu2024halueval}. There are a variety of approaches to improve the ``groundedness'' of LLMs \citep{dhuliawala2023chain, tonmoy2024comprehensive,zhang2023siren}, and for LLMs to delegate deterministic and well-defined tasks such as fact retrieval and arithmetic calculation to authoritative and provably-correct tools \citep{schick2024toolformer}. We can also design ``co-audit'' interfaces that help users spot and fix errors post-generation \citep{gordon2023co}. In short, the types of errors we call hallucinations are gaps we know how to plug.

A different, more subtle, but more enduring challenge is the way in which generative AI influences work where the answers are not a matter of truth and fact. There is seldom a ``right'' way to express an idea, summarise a document, design a presentation, write a piece of code, compose an email, or design a spreadsheet. Some solutions may be better than others, but even to judge them as such requires a complex qualitative evaluation of several factors. Generative AI might generate a perfectly ``correct'' summary of a document containing no factual errors. But what has it highlighted, and what has it omitted? Is the tone internally consistent, and consistent with the rest of the document? Is it comprehensible to the intended audience? These and innumerable other factors are considered in every knowledge workflow.

The human tendency is to ``satisfice'' \citep{artinger2022satisficing}, searching for solutions that merely meet a minimum aspirational threshold rather than optimal solutions. The likelihood of accepting AI-generated output if it contains no obvious errors is high. Studies have found a tendency to offload cognitive tasks to technology, and that individuals compensate for an unwillingness to engage in effortful reasoning by relying on technology \citep{barr2015brain}. Multiple studies have noted a \emph{mechanised convergence} effect \citep{sarkar2023aiknowledgework} of AI tools on knowledge work, i.e., that when using AI tools, people produce work that is more homogenous and less diverse \citep{arnold2020predictive, dellacqua2023navigating,anderson2024homogenization}. Interaction with opinionated models causes users to shift their own opinions \citep{jakesch2023opinionated}. All of this builds towards a picture that suggests that users have a strong tendency to accept AI output without making meaningful adjustments based on personal critical judgment. Generative AI tools, in producing satisficing answers, have the potential to cause the short-circuiting of these considerations at scale, thus causing knowledge work to go on autopilot.\footnote{As \citet{sellen2023rise} note, the introduction of autopilots in aviation was fraught with challenges of vigilance, takeover, de-skilling, and bias, and inappropriate design decisions and trade-offs led repeatedly to adverse consequences.}



The most important challenge for generative AI, then, is not hallucinations but \emph{critical thinking}. Critical thinking is not a new skill for knowledge work, but it is one that gains particular importance with generative AI \citep{sarkar2024challenge}. As AI handles more aspects of the material production of work, the role of the human shifts to \emph{critical integration} \citep{sarkar2023aiknowledgework}. Critical integration involves deciding whether, when, and how to apply generative AI to a workflow, as well as the evaluation and refinement of individual instances of AI output. Several aspects of a critical integration workflow introduce new cognitive and metacognitive demands on users \citep{tankelevitch2023metacognitive}, such as developing effective mental models for prompting strategies.

\subsubsection*{Critical Thinking and Spreadsheets} We have discussed critical thinking for knowledge work in general. To help exemplify our ideas, our prototype (described in Section~\ref{sec:prototype}) focuses on critical thinking of the kind invoked when working with spreadsheets, and in particular in the activity of data-driven shortlisting (described in Section~\ref{sec:shortlisting_motivation}). Spreadsheets make a particularly interesting case study because the scope of risk and error in spreadsheets has a long history of popular and scholarly scrutiny \citep{panko1998we,powell2008critical}, which makes them unique amongst application types that have extremely wide use across domains, professions, and organisations. The impacts of spreadsheet-related errors are huge and continuous, as carefully documented by EuSprIG.\footnote{\url{https://eusprig.org/research-info/horror-stories/}. Last accessed: 23 April 2024.} Errors have numerous sources \citep{panko2010revising,rajalingham2008classification}: incorrect formulas, incorrect data, issues with data structuring \citep{chalhoub2022freedom}, errors in comprehension \citep{ragavan2021comprehension}, unit-related errors \citep{williams2020units}, and mistakes in copy-paste reuse \citep{joharizadeh2020gridlets}, are but a few among many. Adding generative AI tools that can produce formulas, charts, and other types of spreadsheet content from natural language queries, but are subject to hallucinations, would seem to exacerbate the problem by introducing yet another potential source of error. On the other hand, as previously mentioned, hallucinations are not an insurmountable challenge, and moreover, generative AI might even resolve traditional sources of error by improving formula authoring, data structuring, debugging and comprehension, etc., but replacing them with new considerations for awareness, agency, self-efficacy, and attention investment, in what has been termed the \emph{generative shift} \citep{sarkar2023eupgenai}.

Instead, the focus on the challenges of generative AI in spreadsheets must turn to critical thinking. Spreadsheets are \emph{praxisware} \citep{sarkar2023should}; complex software whose feature set is rich and broad, which is used intensively by individuals over long periods of time (often entire careers), which require deep investment in learning and skill development, which can form a core part of one's professional identity and sense of organisational value, and around which informal and formal communities of users can rally and exchange information and ideas. Working with spreadsheets requires multiple, deeply inter-related types of expertise \citep{sarkar2016phd}: \emph{representational} expertise in the spreadsheet user interface, formula language, etc.; \emph{process} expertise in data analysis, analytical formulas and methods, their applications and limitations; and \emph{domain} expertise about the data in their spreadsheets, its correspondence to real-world quantities, knowledge about its provenance and intended use, its limitations, etc. Spreadsheets are also sites of specialised types of critical thinking such as sensemaking \citep{pirolli2005sensemaking, drosos2024rubberduck}, and analytical and statistical thinking \citep{wild1999statistical}. And finally, the user surface of spreadsheets is enormous: the market leaders Microsoft Excel and Google Sheets are estimated to have as many as 1.5 billion and 900 million users, respectively.\footnote{e.g., \url{https://earthweb.com/excel-users/}, \url{https://earthweb.com/google-sheets-users/}. Last accessed: 23 April 2024. User counts are impossible to independently verify, and depend on the method of calculation. These figures should only be taken as an indication of the order of magnitude.} For these reasons, spreadsheets make a practical and impactful testbed from which to begin exploring interventions for critical thinking that may then generalise to more types of knowledge work.


The rest of the paper is organised as follows:
\begin{enumerate}

\item We give a brief overview of the theory of critical thinking and how it has been operationalised in education, and how decision-making can be supported by generative AI (though not always with an explicit focus on critical thinking) (Section~\ref{sec:literature}).

\item Next, we explain the task of interest in this paper: \emph{shortlisting} (Section~\ref{sec:shortlisting_motivation}). Shortlisting is the process of selecting a small set of items from a larger set. It is a ubiquitous class of sensemaking task, and is scoped narrowly enough for detailed study, yet complex enough to involve a wide variety of critical thinking behaviours.

\item We then describe a prototype for \emph{critical shortlisting} (Section~\ref{sec:prototype}) that exemplifies some design ideas about how a tool for shortlisting could use generative AI to induce critical thinking in the user. The core of the prototype are \emph{provocations}: short textual ``reflections'' generated via a large language model that aim to critique, subvert, highlight bias, weaknesses, limitations, and alternatives to the factors being considered for shortlisting. Our designerly intention is that through exposure to such provocations, the user's level of critical reflection during a task is heightened.

\item Finally, we describe a research agenda for critical thinking in generative AI-assisted knowledge work (Section~\ref{sec:research_agenda}).

\end{enumerate}


\subsection{Critical Thinking}
\label{sec:literature}




\subsubsection*{Frameworks}

``Critical thinking'' is a widely-used term with no definitional consensus. Indeed, the term is contested and debated and even its utility as a concept has been questioned. The colloquial use of the term has not helped. Dictionary entries, perhaps as a metacognitive challenge, offer circular descriptions such as \emph{``the act or practice of thinking critically (as by applying reason and questioning assumptions) in order to solve problems, evaluate information, discern biases, etc.''}\footnote{\url{https://www.merriam-webster.com/dictionary/critical\%20thinking}. Last accessed: 23 April 2024.}

An influential framework which can serve as a starting point is the hierarchical taxonomy developed by \citet{bloom1956taxonomy}, which characterises student learning objectives starting with simple recall of knowledge and leading to analysis (distinguishing and relating ideas), synthesis (combining ideas into a novel form) and evaluation (appraising ideas according to some criteria). The latter three are often combined to serve as a definition for critical thinking, though it is worth noting that the taxonomy distinguishes between creative thinking (generating novel ideas) and critical thinking (assessing ideas according to specific criteria), both of which are given equal status at the top of the hierarchy \citep{huitt2011bloom}.

Applying the Delphi method of expert consensus \citep{facione1990critical}, Facione developed a theory of critical thinking \citep{facione2011critical} arriving at very similar categories (though not hierarchically ordered) such as interpretation, analysis, inference, and evaluation. Additionally, they suggested that the application of critical thinking skills is substantially dependent upon a deeper psychological \emph{disposition} \citep{facione1995disposition}, which must also be trained in addition to training critical thinking skills. The disposition is conceived of a as a set of seven intellectual attitudes, which bear a similarity to personality traits (Inquisitiveness, Open-mindedness, Systematicity, Analyticity, Truth-seeking, Self-confidence and Maturity).

Another term for critical thinking is reflective thinking (though not all scholars use the terms interchangeably). John Dewey, widely considered the progenitor of the modern concept of critical thinking, originally used the term reflective thinking \citep{dewey1910howwethink}. Nguyen et al. propose a five component model of reflective thinking \citep{nguyen2014reflection}, defining it (in approximate terms) as iteratively engaging with one's own thoughts and actions and their background conceptual frames with a view to changing them. The reflective judgment model \citep{king1997reflective} describes how epistemic assumptions develop and affect an individual's subsequent problem solving. It posits multiple stages of knowing, from stage one (\emph{``concrete, single-category belief systems'')} to stage seven (\emph{``knowing is uncertain [... but nonetheless ...] epistemically justified claims can be made''}). At each subsequent stage the individual develops in their ability to reflect on, judge, and critique knowledge. It is assumed that progression through the stages is sequential, and proposed that by identifying which stage a particular student is in, an educator may help them progress to the next stage. Particular types of argumentation practice and exercises in critical thinking have been developed to help individuals at different stages. In particular the use of \emph{ill-structured} problems is key for developing reflective thinking. Data analysis problems are a convenient fit, especially at the analysis planning stage, because they are extremely ill-structured \citep{sarkar2016constructivist, sarkar2016phd}.

Similar to the reflective judgment model, the Paul-Elder model of critical thinking (sometimes simply the ``Paul model'') also posits a staged model of development \citep{paul1997california}, and advocates for learners to develop structural thinking, standards of intellectual rigour, and traits (attitudes) towards knowledge (such as humility). Staged models of critical thinking exhibit the common property that as learners develop, they not only become increasingly critical of knowledge content but also the processes of generating knowledge and knowledge discourse. This has been conceptualised by \citet{wilson2016teaching} as a \emph{metacognitive} skill, which can be taught through classroom exercises.

Many frameworks that aim to help learners develop critical thinking skills focus on evaluating claims and arguments (deriving from a long preoccupation in Western philosophy with valid forms of argument that can be traced to Aristotle). For example, the Toulmin model of argumentation \citep{kneupper1978teaching} decomposes an argument into constituent elements such as data, warrants, backing, qualifiers, and claims, and exposes their relationships diagrammatically. Similarly, the diagrammatic technique of argument mapping \citep{davies2011concept} decomposes an argument and organises claims, objections, and supporting evidence into a hierarchy which helps assess the overall quality of an argument and identify weaknesses.


However, Willingham questions whether it is possible to consistently teach critical thinking \citep{willingham2008critical}. Drawing from a diverse range of studies with students and meta-analyses thereof, he concludes that \emph{``critical thinking [...] is not a skill''} in the sense that it cannot be distilled into a set of learnable practices that can be applied regardless of context. However, he notes that \emph{``there are metacognitive strategies that, once learned, make critical thinking more likely [...  and] the ability to think critically [...] depends on domain knowledge and practice.''} Examples of Willingham's metacognitive strategies include ``look at both sides of the issue'' and ``look for a problem's deep structure''.

These are merely a sample of some of the key frameworks and ideas in critical thinking. In fact, critical thinking is a highly contested term and many alternative definitions and frameworks have been proposed. An overview is given by \citet{mulnix2012thinking}, who further suggests that despite all of these differences, the various definitions of critical thinking can all be united by (or are consequents of) the foundational skill of \emph{``recognizing the inferential connections that hold between statements''}. Deanna Kuhn similarly posits \emph{``the ability to recognize the possible falsehood of a theory and the identification of evidence capable of disconfirming it [... and more broadly ...] the ability to justify what one claims to be true''} \citep{kuhn1993connecting}.

The succinct definitions proposed by Mulnix and Deanna Kuhn are attractive baselines which informed our work. However, it is worth noting that these definitions of critical thinking adhere to a distinctly Popperian worldview \citep{popper2005logic}, as evidenced by Deanna Kuhn's emphasis on falsificationism, and Mulnix' objectivist description of evidence (\emph{``it is a matter of objective fact whether one statement evidentially supports another''} \citep{mulnix2012thinking}). On the contrary, as other philosophers of science (notably Thomas Kuhn and Paul Feyerabend) have pointed out, it is not a matter of objective fact whether one statement supports another; statements or facts are \emph{theory-laden} and can only be interpreted as evidence for or against other statements against the background of some necessarily subjective theories \citep{kuhn1962structure, feyerabend1975against}. An alternative synthesis of multiple critical thinking frameworks, attempting to integrate the concepts of reflective judgment and metacognition, that does not attempt to offer a reductionist or objectivist foundation for critical thinking, is given by Dwyer et al. \citep{dwyer2014integrated}.

\subsubsection*{Measures}
\label{sec:ct_measures}

Ennis considers how critical thinking can be assessed \citep{ennis1993critical}, suggesting that simple multiple choice questions are insufficient, and suggests more evaluation intensive alternatives such as adding justifications to multiple choice questions, structured essays, and qualitative whole-portfolio assessment.

But even within a tightly circumscribed domain, such as nursing, there is a wide variety of opinions on how critical thinking should be taught and evaluated, including task observation, qualitative assessment by clients (patients), discussion with peer interaction, and examination by multiple choice questions or essays \citep{PAUL20141357}.

A widely used questionnaire for assessing critical and reflective thinking was developed by \citet{kember2000development}. The instrument consists of sixteen 5-point Likert items (agree-disagree), measuring 4 categories of critical thinking (habitual action, understanding, reflection, and critical reflection). Similarly, \citet{kobylarek2022critical} have recently developed a 25-item questionnaire that aims to directly quantify critical thinking skill at each level of the Bloom taxonomy. These questionnaires are completed by thinkers themselves (rather than by assessors) so it is a self-judged measure of critical thinking ability. Conversely, \citet{kember2008four} have also shown how their 4-category scheme can be used as an assessment tool by educators to assess the level of critical thinking exhibited in written work.

\citet{facione1994critical} show how frameworks for critical thinking can be refined into a domain-specific questionnaire (in this case, clinical nursing practice). Their inventory consists of seven subscales of critical thinking disposition relevant to nursing, such as inquisitiveness, systematicity, open-mindedness, and self-confidence. Similar work has been recently conducted by \citet{zuriguel2017development, zuriguel2022nursing}.

\subsubsection*{Design Research for Critical Thinking}

\citet{sun2017critical} studied how a web application for developing structured arguments helps pairs of learners think critically together. They found that horizontally juxtaposing opposing arguments (e.g., pros and cons) helped learners compare arguments. They found that learners communicated with their partners to clarify the wording, validity, and content of their arguments, as well as edited each other's arguments directly. Similarly, \citet{TSAI2015187} evaluate an online visualisation tool that helps learners map arguments onto the aforementioned Toulmin model of argumentation, finding that it can lead to a statistically significant reduction in pseudoscientific beliefs (such as in horoscopes, and magnetic therapy).


\citet{lee2023fostering} studied how engaging with an art exhibition can foster critical thinking about AI in youth. They found that such engagements can facilitate active discussion and sharing of opinions with a peer group, contrary to previous findings relating youth with passivity and a reluctance to express opinions. These tension or conflict-filled discussions were central to development of criticality. They also found that the engagement broadened users' perceptions of AI away from mental pictures of specific tools (e.g., ``Siri'') to considering many wider applications of AI and their attendant ethical challenges. Nonetheless, they did not observe evidence sufficient to conclude that the exhibition had resulted in a long-term improvement in critical thinking competency.

\citet{holzer2015mobile} developed a mobile application for improving users' critical thinking about online misinformation based on Carl Sagan's ``Baloney Detection Kit''. The kit includes questions to encourage thinking about the quality of evidence for a claim, potential alternative claims, and potential fallacies in the argument. However, the effectiveness of the application for developing critical thinking skills was not evaluated. A later study \citep{holzer2018debate} evaluated participants' beliefs before and after a critical thinking lesson involving the Baloney Detection Kit, finding a statistically significant belief change in the direction best supported by the evidence.




\citet{reicherts2022chatbots} study the possibility of introducing proactive chatbots which intervene when the user is about to make an important decision, and encourage them to think carefully about their assumptions and what they intend to do. They evaluate this in the context of stock investment decisions. Participants in the study reported that \emph{``probing interactions could enhance their own decision-making by preventing certain impulsive actions or inconsistencies in decisions''}.


\citet{danry2023ask} conducted an experiment where participants were shown socially divisive statements accompanied either by no additional information, or one of two types of AI-generated information: a statement explaining the AI evaluation of the argument, or a question that prompts the user to evaluate the argument in some way. They found that the questioning format led to a significant improvement in human logical discernment over the other two formats. This reflects research in social psychology that has found that when students are asked to write a counter-attitudinal essay (i.e., to make an argument that contradicts their own personal beliefs), students who self-generate the arguments in this essay (as opposed to those who are asked to summarize arguments made by others) are more likely to change their attitudes and subsequent behaviours \citep{miller2001counter}.



\citet{barr2015brain} found that people may offload cognitive tasks to technology: \emph{``individuals who are relatively less willing and/or able to engage effortful reasoning processes may compensate by relying on the internet through their Smartphones. Across three studies, we find that those who think more intuitively and less analytically when given reasoning problems were more likely to rely on their Smartphones [...] for information''}. Moreover, \citet{saade2012critical} found that learners perceived the interactive components of an e-learning platform (such as quizzes and exercises) to contribute more to the development of their critical thinking than static digitalised resources (such as online textbooks). This suggest a tremendous opportunity, as well as a responsibility, for system designers to provide support for critical and creative thinking through technology.



\subsection{Shortlisting}
\label{sec:shortlisting_motivation}

Shortlisting, the process of selecting a small set of items from a larger set, is a ubiquitous task in many domains. For example: A hiring manager shortlists job candidates for interviews from a large pool of applicants. A charity grant fund manager shortlists charities to receive donations from a large set of grant applications. A house hunter shortlists apartments to view from a large list of search results. A software product manager shortlists features to implement from a large list of team ideas and user requests. While common and important, shortlisting tasks are sufficiently small in scope (compared to open-ended exploratory data analysis) to be suitable for detailed and controlled study, yet still sufficiently interesting, diverse, and ecologically valid to provide relevant design insights for critical thinking.


Despite its prevalence, shortlisting as a form of data analysis is understudied and underserved by current tools. Artificial intelligence (AI) language models could potentially aid in shortlisting by suggesting relevant factors and evaluating how well each candidate aligns with those factors. But shortlisting is a complex task encompassing a heterogenous set of cognitive and critical thinking activities. It involves the application of decision criteria to candidates, while simultaneously evaluating the relative importance and definition of those very criteria. It requires synthesizing both qualitative and quantitative variables into an overall assessment for each item under consideration. Tasks such as selecting job candidates, grant recipients, or product features are inherently ambiguous and lack clear-cut solutions.




Traditionally, we have viewed AI language models as metaphorical assistants that enhance our productivity. While mitigating errors and hallucinations remains crucial, as AI expands into more complex domains lacking clear-cut solutions, the primary challenge shifts from task completion to task definition, from solving the problem to formulating the problem. 




In these ambiguous scenarios, the AI has the potential to serve as more than an assistant -- as a provocateur: an entity that fosters critical thinking \citep{sarkar2024challenge}. In the following section (Section~\ref{sec:prototype}) we describe a first prototype that attempts to integrate generative AI into the process of shortlisting, as an assistant but also as a provocateur. The core manifestations of the AI provocateur in our prototype are ``provocations'', short text messages which aim to critique and identify limitations, biases, risks, weaknesses, and opportunities for improvement in the AI assistant's suggestions. Provocations represent a more nuanced and critical evolution of the disclaimer text often accompanying AI models, such as ``AI can make mistakes.'' The user can then choose whether to engage with the AI's self-critique or disregard it.

Shortlisting presents a suitable test bed for studying the provocateur role of AI for several reasons:

\begin{itemize}
    \item Diverse decision factors: Shortlisting tasks involve considering a wide range of factors, both quantitative (e.g., job applicant scores, grant proposal budgets) and qualitative (e.g., cultural fit, alignment with organizational values). This diversity of factors mirrors the complexity of real-world decision-making processes and provides multiple opportunities for AI to generate thought-provoking critiques and alternative perspectives.
    \item Subjective evaluation: The absence of clear-cut solutions in shortlisting tasks necessitates subjective evaluation by human decision-makers. AI systems can be designed to highlight the subjectivity inherent in their suggestions, prompting users to critically examine their own biases and assumptions.
    \item Iterative decision-making: Shortlisting is often an iterative process, with decision-makers refining their criteria and selections over multiple rounds. This iterative nature aligns well with the envisioned role of AI as a provocateur, continuously challenging the user's thinking and fostering ongoing critical evaluation.
\end{itemize}

Studying shortlisting tasks allows the exploration of various interaction modalities and UI elements that facilitate critical thinking. For example, the system could be designed to present contrasting perspectives, highlight potential biases or inconsistencies in the user's decision-making process, or provide interactive visualizations that encourage exploration of different decision factors and their relative weights. These insights could have broader implications for the design of AI-augmented decision support systems across various knowledge work domains.

It is worth noting that when generative AI is applied to complex, open-ended scenarios, its output may be flawed or limited, but not in an easily detectable and rectifiable manner (e.g., a factually incorrect hallucination). However, in cases where there is a clear error, a critical thinking intervention or provocation is the wrong kind of intellectual tool: in such cases we require automated help in finding and fixing such mistakes, i.e., ``co-audit'' \citep{gordon2023co}. We are not primarily concerned with mistakes of this kind.

\section{Critical Shortlisting Prototype}
\label{sec:prototype}

As a first exploration of the idea that AI can promote critical thinking by acting as a provocateur, we developed a prototype in which users can apply AI to help them shortlist a smaller set of items from a larger set. The user uploads a dataset and enters an initial query explaining their shortlisting problem. From there, the system suggests factors that the user might consider, and automatically sorts rows in the dataset by their relative strength according to these factors. Crucially, the system also generates provocations that might cause the user to edit, augment, and discard factors, or simply reflect on the implications of using a particular factor to rank candidate rows. The remainder of this section describes the prototype in detail.

\subsection{Interface Overview}
\label{sec:interface_overview}
This section explains the components of the user interface. Users can load a dataset and enter an initial prompt (Section~\ref{sec:initial_prompt}). The system generates a set of potential factors which can be individually inspected and edited (Section~\ref{sec:factor_cards}). For each factor, a factor analysis is generated with a descriptive overview of the data in the relevant columns and a shortlist specific to that factor (Section~\ref{sec:factor_analysis}). The global dataset and the global system-generated shortlist can be viewed in a single table (Section~\ref{sec:global_shortlist}).

\subsubsection{Dataset Loading and Initial Prompt}
\label{sec:initial_prompt}
\begin{figure}[h]
  \centering
  \includegraphics[width=.45\linewidth]{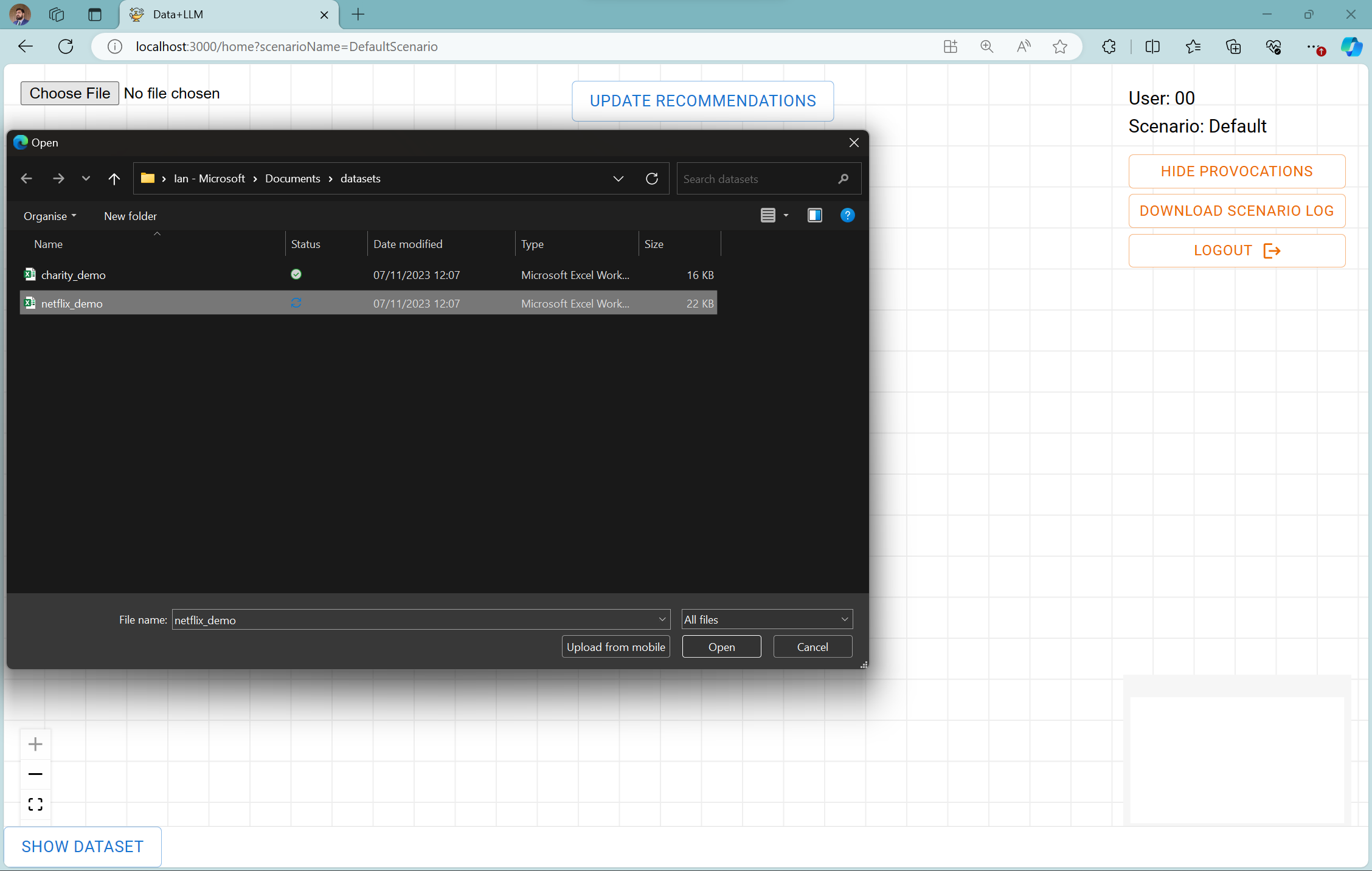}
  \includegraphics[width=.45\linewidth]{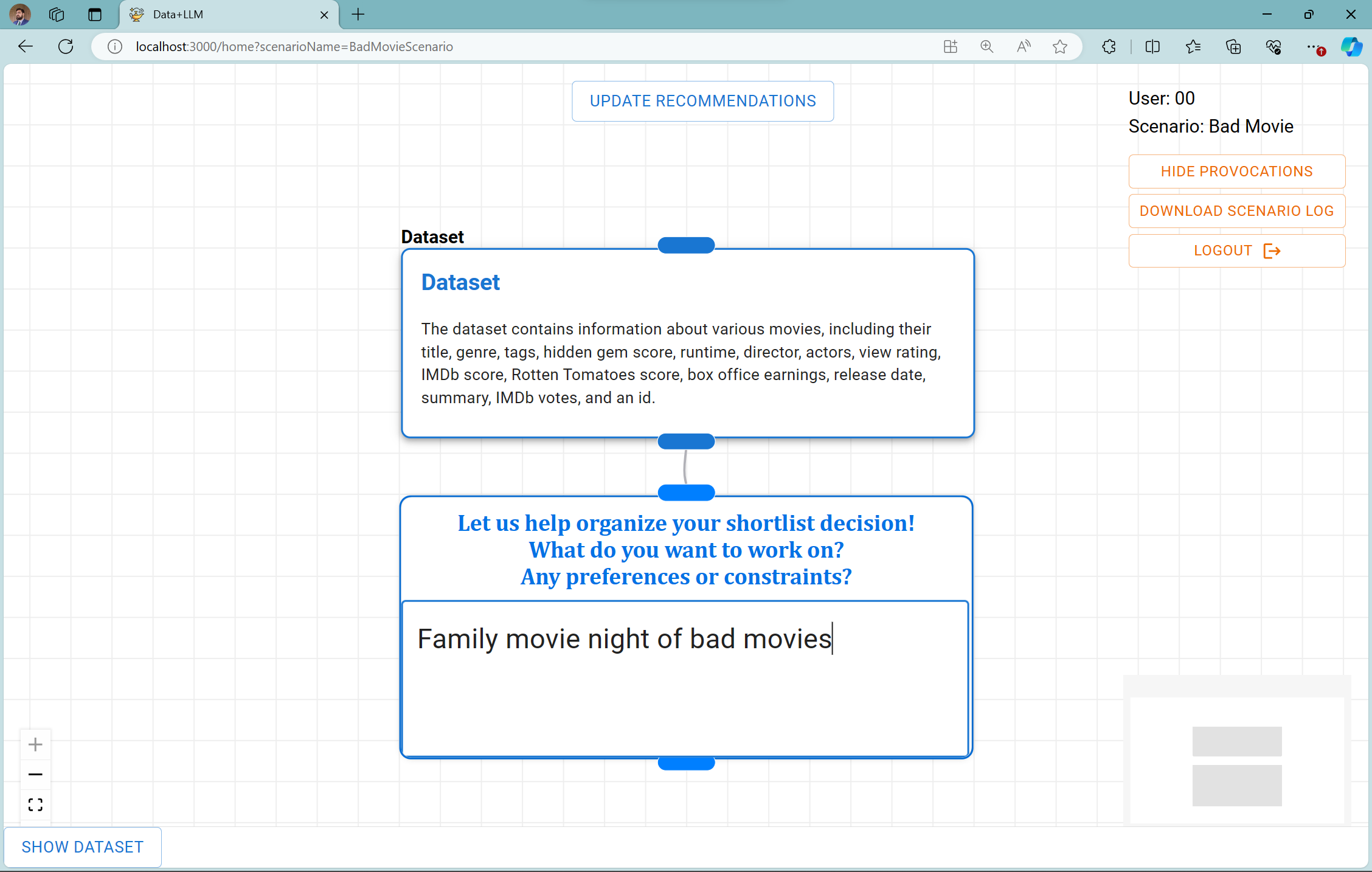}
  \caption{Dataset loading and initial prompt.}
  \label{fig:initial_prompt}
\end{figure}

A dataset in CSV (Comma Separated Values) or Microsoft Excel format can be loaded. The dataset must conform to a well-defined relational schema and contain a header row. Once loaded, the user can enter their initial query, specifying their overall intent and any desiderata for the shortlisting activity. An example is given in Figure~\ref{fig:initial_prompt}.

\subsubsection{Factor Cards}
\label{sec:factor_cards}

\begin{figure}[h]
  \centering
  \includegraphics[width=.45\linewidth]{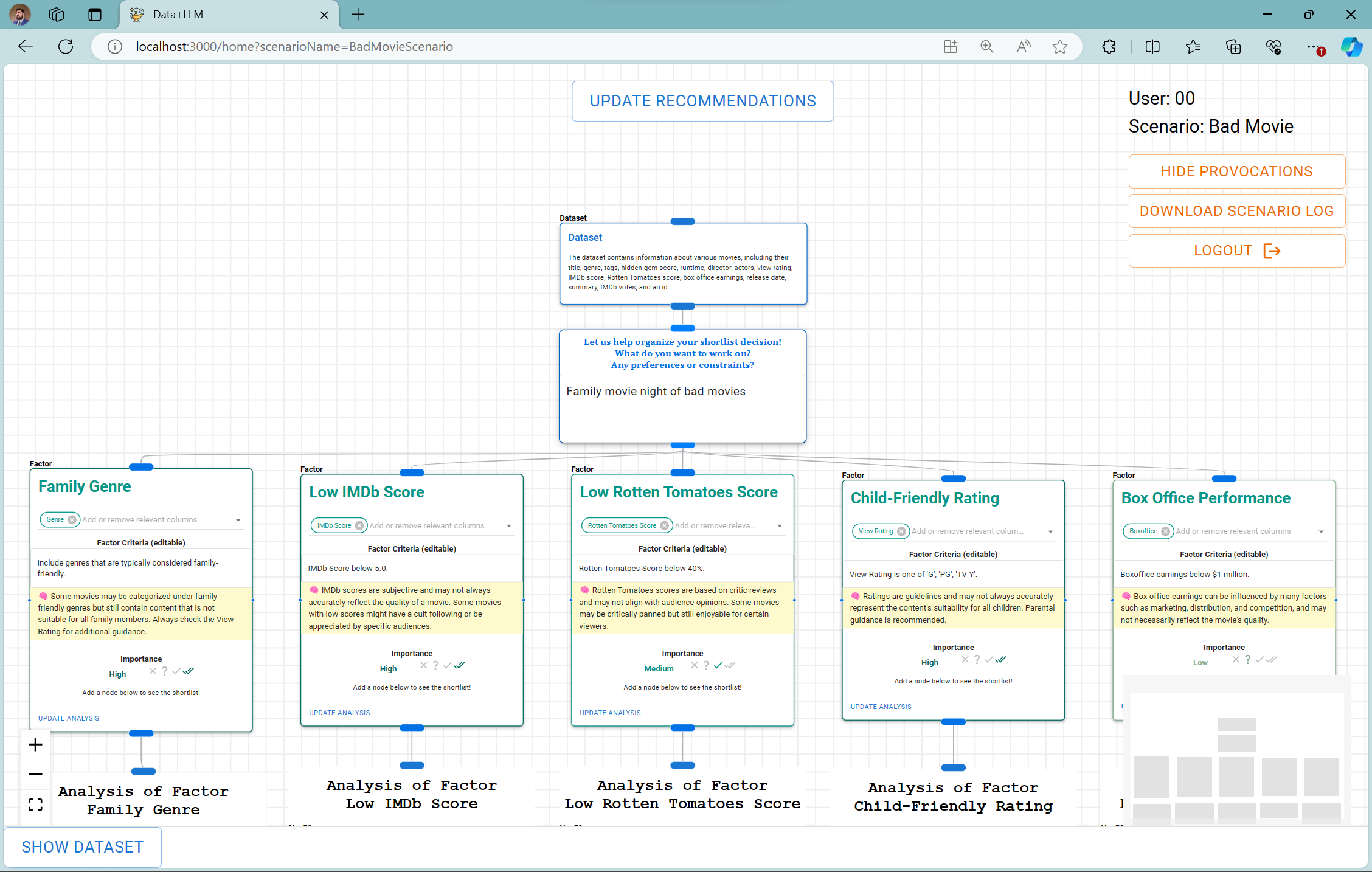}
  \includegraphics[width=.45\linewidth]{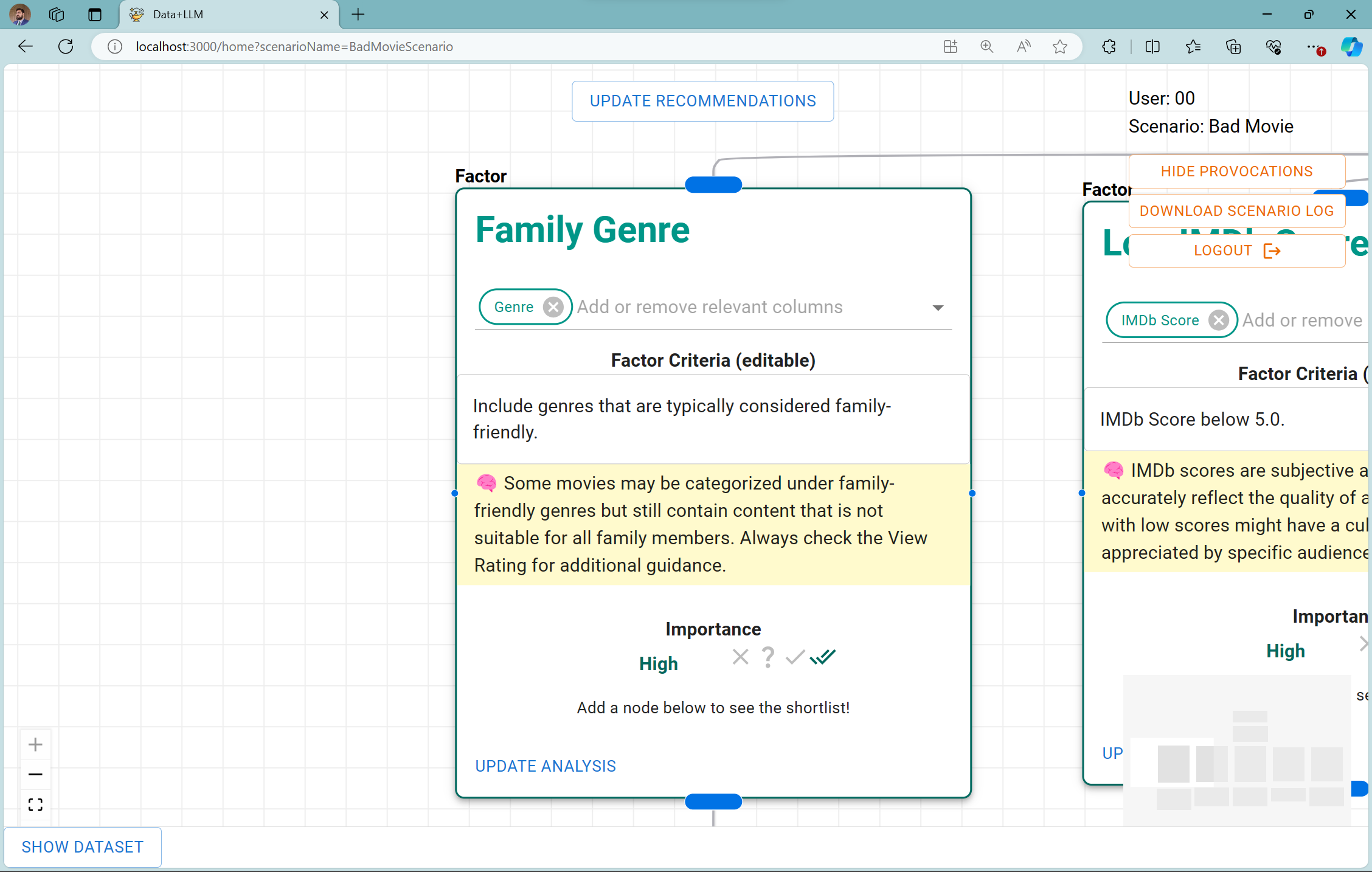}
  \caption{Factor cards}
  \label{fig:factor_cards}
\end{figure}

When a query is entered, the system generates a number of factor cards (Figure~\ref{fig:factor_cards}). The prototype is constrained to generate no more than 5 factor cards to avoid overwhelming the user. Each factor card consists of the following components:

\begin{enumerate}
    \item \textbf{Factor title}: a short descriptive title for the factor.
    \item \textbf{Source columns}: a list of columns from the source dataset which will be used in evaluating each candidate row with respect to the factor. This can be edited to add or remove columns. The system can also generate factor suggestions for which the system cannot determine a suitable source column (it may even be that the data required does not exist in the dataset), and in such cases the list of source columns is left empty and the factor cannot be applied until the user has indicated a suitable source column or added the necessary data to the dataset.
    \item \textbf{Factor criteria}: a description of the factor representing desiderata for candidate rows. This description is used as part of the prompt for a language model to evaluate each candidate row. This can be edited to contain any arbitrary text.
    \item \textbf{Factor provocation}: a system-generated critique of the factor which may highlight potential limitations and suggest improvements or alternatives.
    \item \textbf{Factor importance}: a set of buttons allowing the user to customise the weight that this factor is given in computing the global shortlist. To produce the global shortlist, each row is scored according to a sum of their factor scores weighted by their factor importance.
    \item \textbf{Update analysis button}: when clicked, this button applies the model to update the factor analysis (Section~\ref{sec:factor_analysis}) and the global shortlist (Section~\ref{sec:global_shortlist}).
\end{enumerate}

Factor cards can be deleted. New factor cards can be created by clicking and dragging the handle below the query box to ``spawn'' a new card.

\subsubsection{Factor Analysis}
\label{sec:factor_analysis}

\begin{figure}[h]
  \centering
  \includegraphics[width=.45\linewidth]{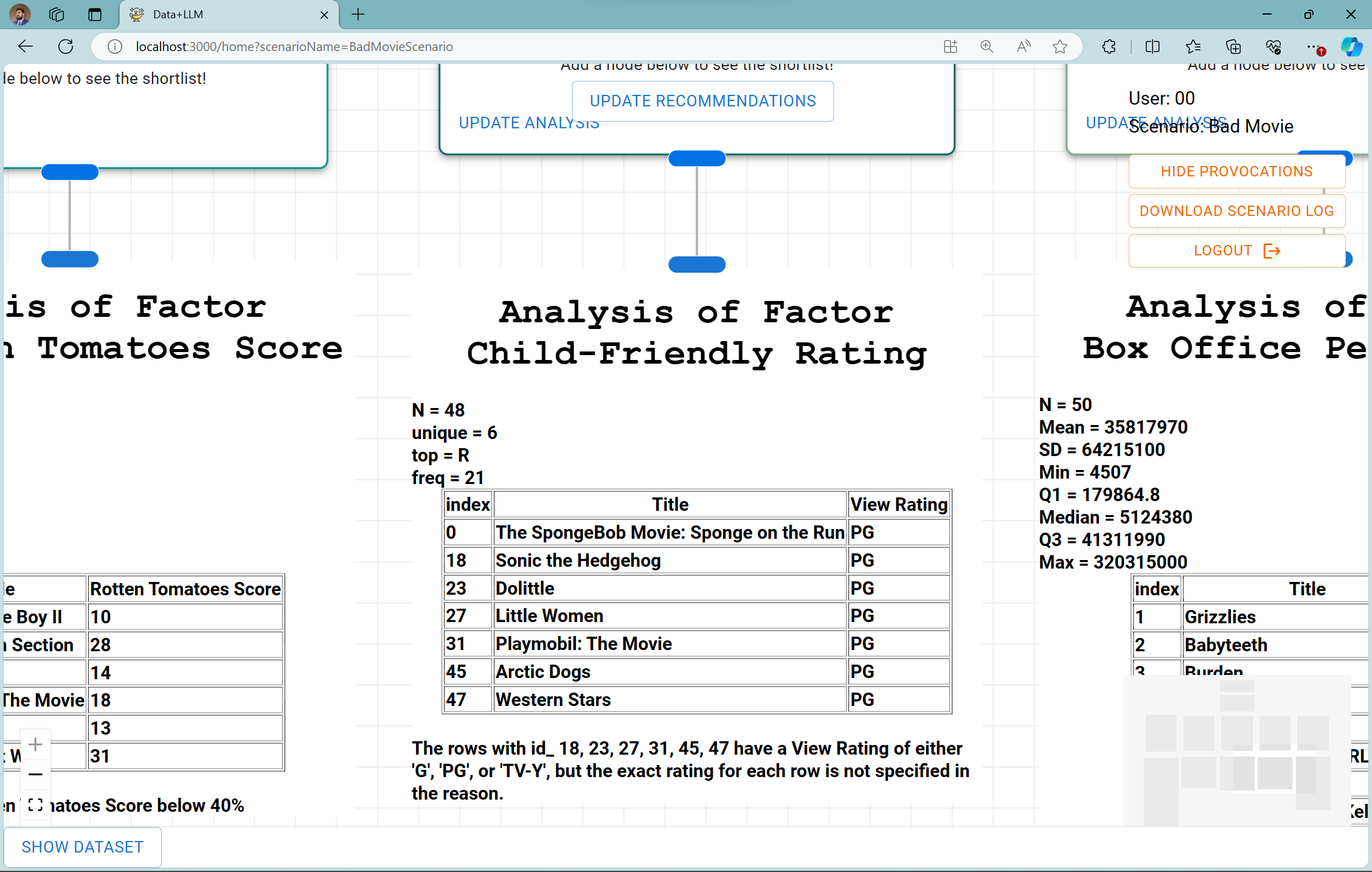}
  \includegraphics[width=.45\linewidth]{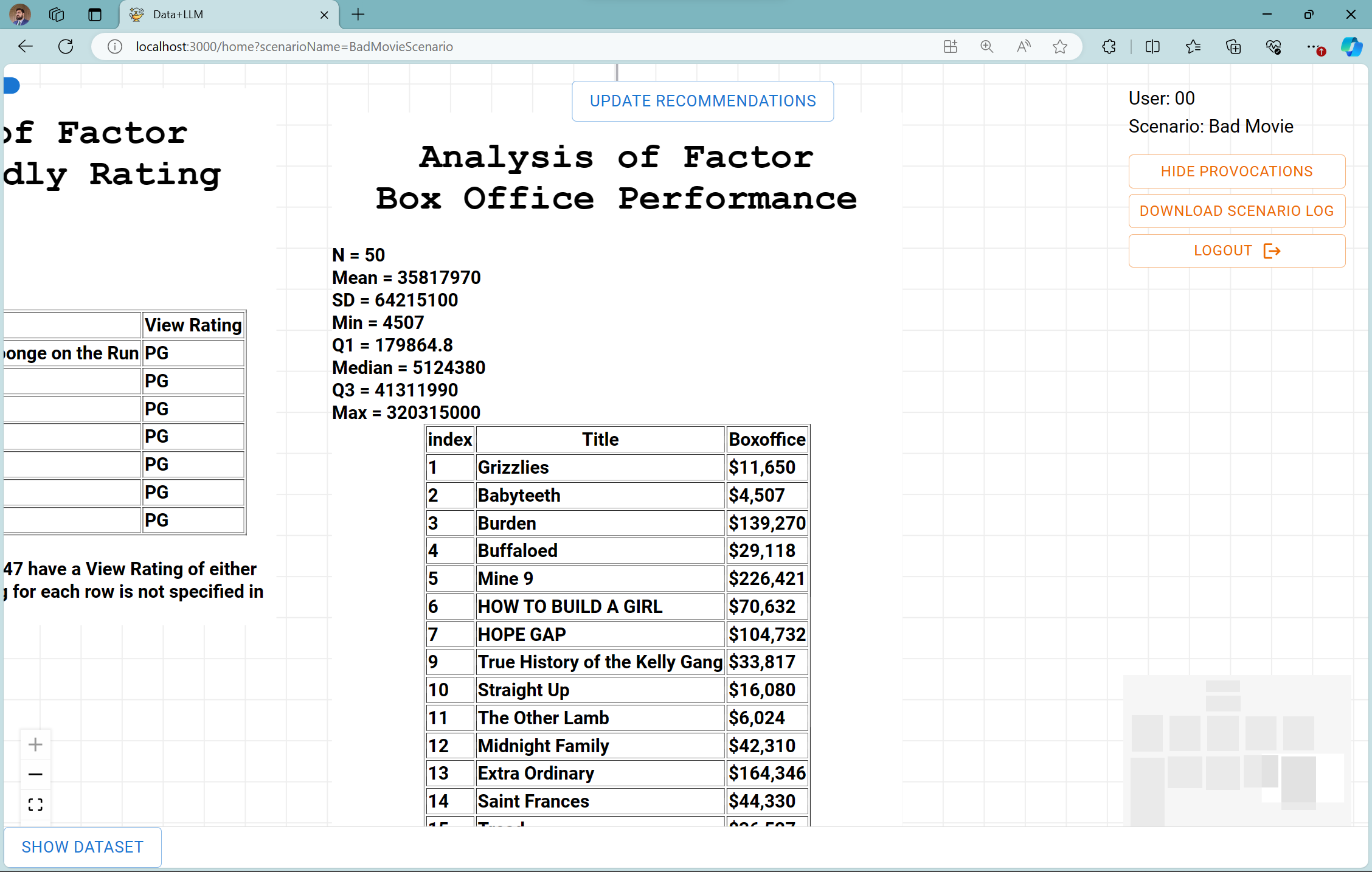}
  \caption{Factor analysis}
  \label{fig:factor_analysis}
\end{figure}

Below each factor card is a factor analysis (Figure~\ref{fig:factor_analysis}) that is produced and updated when the ``update analysis'' button on the corresponding card is clicked. The factor analysis provides a summary of the data in the column. If the column contains numerical values, the factor analysis will show summary statistics such as the mean, median, minimum, maximum, standard deviation, etc. If the column contains text, the factor analysis shows frequently occurring values.


Below the data summary, the factor analysis produces a \emph{factor-local} shortlist, i.e., a shortlist of all the candidate rows in the dataset that score highly according to that single factor, ignoring all other factors. This allows the user to independently test and refine each factor, and allows for comparative analysis between factor-local and global shortlists (e.g., it is helpful to consider why a candidate row appears in a factor-local list but not the global list, or vice versa).

\subsubsection{Global List View}
\label{sec:global_shortlist}

\begin{figure}[h]
  \centering
  \includegraphics[width=.95\linewidth]{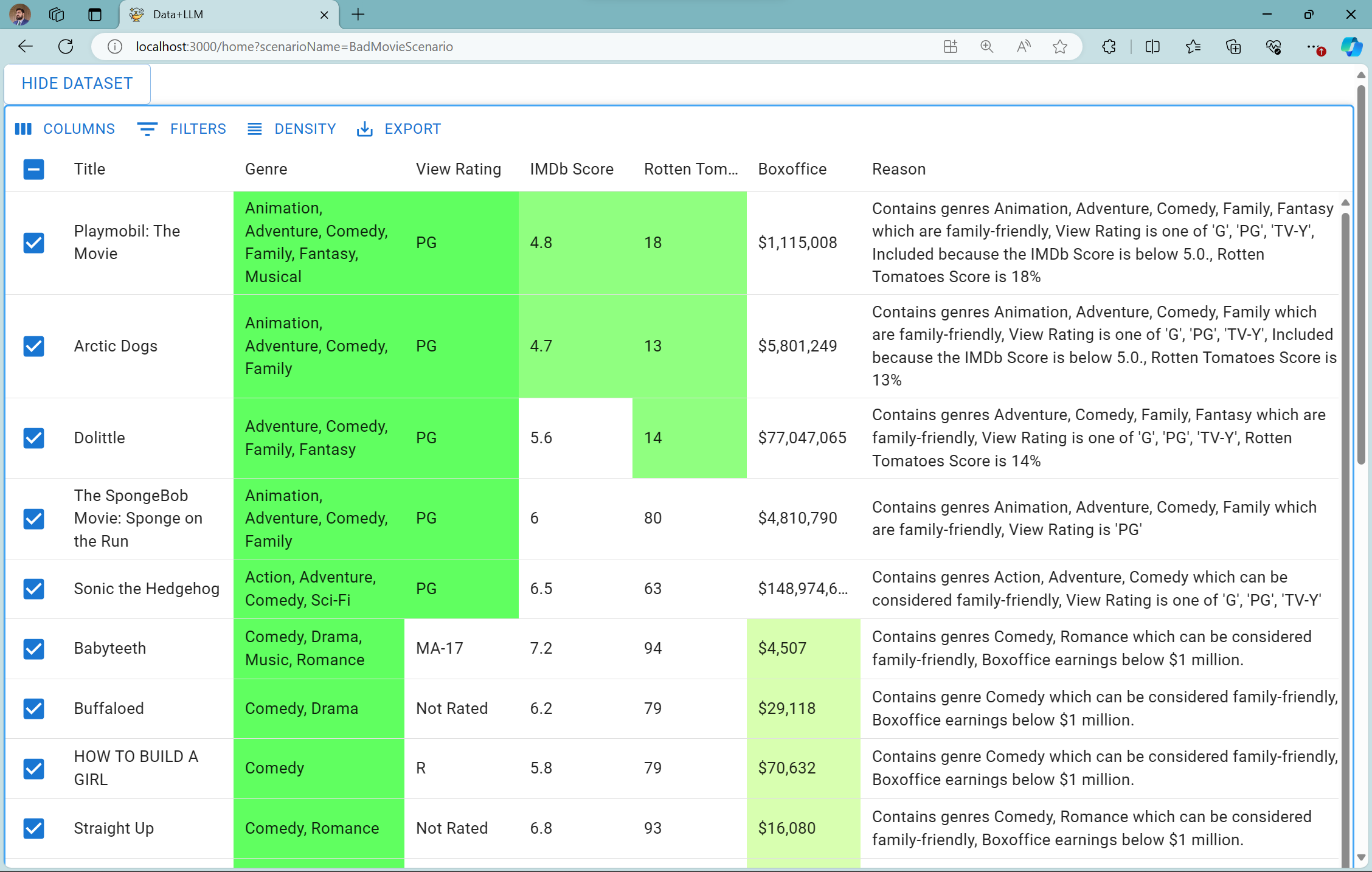}
  \caption{Global list}
  \label{fig:global_list}
\end{figure}

The global list view (Figure~\ref{fig:global_list}) can be accessed from a button at the bottom of the interface. This shows a standard table component containing all the candidate rows from the dataset. When at least one factor analysis has been applied, or the ``update recommendations'' button is pressed, this list is re-sorted to place the shortlisted candidate rows at the top, ranked by their weighted factor scores.




To calculate weighted factor scores, first, each factor $f$ is given a weight $w_f$ according to its importance; i.e. factors with High importance are assigned $w_f = 1$, for Medium importance $w_f = 0.66$, etc. The weighted factor score for each row is the sum of all $w_f$ for which the row is in $f$'s factor-local shortlist. This shortlisting method allows us to reduce multiple criteria and weights into a single dimension along which rows can be ranked. However, our choice of this particular weighting and ranking method involves several ad-hoc decisions that were largely pragmatic decisions for implementation simplicity. If future work is interested in the shortlisting activity \emph{per se}, this is an important area to revisit. For our purposes, we are more interested in the prototype as a testbed for critical thinking and so a detailed investigation of ranking methods is out of scope.


Within each shortlisted row, columns which participate in a factor are coloured green to indicate that they scored highly for a factor. Different shades of green are used to indicate different factor weights; higher factor weights are indicated with higher saturation.

The rightmost ``Reason'' column contains a system-generated summary of how each candidate row was scored according to the factors. This allows the user to check whether the system is correctly applying each factor to each candidate row.


\subsection{Example User Scenario: Choosing Movies for Family Movie Night}
For an example workflow with this prototype, consider Clara who wants organise a movie night for her family. Clara first uploads a dataset of current Netflix movies she downloaded online (Figure~\ref{fig:initial_prompt}~left). The prototype generates a short description of the dataset and provides a prompt window where Clara can provide preferences and constraints she has for her shortlist to the prototype in natural language. Clara adds that she wants to make shortlist of movies for a \emph{``Family movie night of bad movies''} since she and her family love making fun of bad movies together (Figure~\ref{fig:initial_prompt}~right).

The prototype generates a list of factors based on Clara's input for her to consider (Figure~\ref{fig:factor_cards}~left). Clara inspects these factors and views the title, description of the criteria in how the factor will be applied, and reads the provocation text in yellow to see if there are any changes to factor criteria or importance, or even adding new factors, that she might want to do as part of her shortlisting activity (Figure~\ref{fig:factor_cards}~right).

Clara investigates how each factor applies to her data by clicking ``Update Analysis''  (Figure~\ref{fig:factor_cards}~left and right) which generates a factor-local analysis based on the criteria selected and gives Clara a preview of the rows selected based solely on this factor card (Figure~\ref{fig:factor_analysis}~left and right).

Satisfied with the factor cards, Clara selects \emph{``Show Dataset''} to view the global shortlist which has highlighted various cells based on the factors investigated by Clara, which has sorted the rows with the most matching factors (based on importance weighting) to the top of the shortlist (Figure~\ref{fig:global_list}). 

Clara inspects the shortlist and sees that even though the factor criterion for \emph{``Box office performance''} has not been met by the top three movies, they satisfy most or all of the other more important factors. Clara confirms this by reading the generated \emph{``Reason''} column which explains how each row meets the criteria specified (Figure~\ref{fig:global_list}). Having finalised the shortlist of movies for her movie night, Clara opens Netflix, queues up the top three movies in the shortlist, and gathers her family into the living room for their movie night.

\subsection{Prototype Status and Next Steps}

Our prototype is a work in progress and represents one example of our evolving ideas on critical thinking with generative AI. Details of the prototype implementation are given in Appendix~\ref{apx:implementation}. The intention is that this particular prototype can serve as the first point in a larger space of critical thinking interventions. Future work (a plan for which is sketched in Section~\ref{sec:research_agenda}) will involve creating a range of prototypes occupying different points in this space, forming a portfolio of design techniques that allow us to better understand how to design such interventions. Our next step is to evaluate the prototype in a user study where participants attempt shortlisting tasks. We expect to learn whether provocations can have a meaningful effect on participants' critical thinking during shortlisting. We are interested in developing some empirical basis for understanding which kinds of situations benefit from critical thinking provocations, which don't, and the advantages and disadvantages of the particular form of provocations we are currently producing and displaying.

The prototype is to be understood as a research probe rather than a proposition for an actual product. It has limitations in the size of dataset it can robustly load and display, the speed of response and critique generation, and its spreadsheet capabilities are an extremely narrow subset of those available in commercial spreadsheet software. As such, the aim of our evaluation is not the usability of the system or our particular visual layout \emph{per se}. We are focused narrowly on the effects (or non-effects) of provocations on critical thinking.



\section{Critical Thinking With AI: A Research Agenda}
\label{sec:research_agenda}

Our vision is to provide users with the tools and skills necessary to think critically about AI output and facilitate their ability to integrate it effectively. Clearly, this objective extends beyond spreadsheets. As generative AI is applied to more knowledge work activities and included in more applications, the need for critical thinking applies to each engagement with AI, including the full gamut of generative AI-assisted tasks, such as web search, programming, image generation, document summarization, presentation design, data analysis, robotic process automation, text generation, question answering, translation, etc. Every interaction and every task is an opportunity for users to think critically, learn, and grow.



We are successful in the short term, when, because of our design, the user thinks more deeply, more broadly, more comprehensively, with more self-reflection, and more confidence, i.e., more critically, about AI output in their task.

However, our ultimate success lies in the long term impact we have on users. We are successful in the long term, when, because of our design, the user learns and grows as an individual and professional. The aim is to achieve enduring cognitive benefits that transcend any particular interactions and tasks.







How are we to achieve this objective? The research agenda for critical thinking in knowledge work can broadly be divided into questions of interaction design and questions of technical implementation (acknowledging that these are interdependent and cannot be cleanly separated).

\subsection{Interaction Design Questions}

\subsubsection*{Theoretical Foundations}
A first set of questions seeks to provide theoretic and empirical foundations for the design of critical thinking tools for AI-assisted knowledge work.

As previous work has noted \citep{sarkar2024challenge}, critical thinking support has been studied extensively in the context of education, and a few specific professions, such as history \citep{seixas2004teaching}, nursing \citep{rn2002critical}, and psychology \citep{sternberg2020critical}. However, there is a gap in our understanding of the way in which critical thinking needs arise in many other ordinary workflows for contemporary knowledge work. Understanding and summarizing a document, answering questions from a database, performing data analysis in a spreadsheet, developing a presentation, writing and editing code, reading and responding to email: many activities generalise to multiple professional roles and industries, yet the particular nature of critical thinking in these activities is not well studied. Through interviews or diary studies (for example), we might ask: what are the practical critical thinking needs of knowledge workers? What are the opportunities and challenges for critical thinking in professional knowledge workflows?



Evaluation is another piece of the puzzle. How do we measure critical thinking outcomes? How do we evaluate within-task critical thinking outcomes and internal benefits -- i.e., whether our design is improving the quality of work and user thinking for each specific task they engage in? How do we understand the out-of-task benefits of critical thinking -- i.e., benefits of critical thinking that have other external impacts beyond the immediate task at hand, e.g., improved collaboration with others? In contrast, how do we evaluate internal benefits to the user, e.g., increased self-confidence, feelings of agency and satisfaction? Previous work has developed a number of potential solutions, such as self-reflection questionnaires and expert assessment, and others reviewed in Section~\ref{sec:ct_measures}. It is an open question whether these techniques can be easily imported from their original settings of education and specialised professions into more general knowledge work. Moreover, is there a way to scale or even automate such evaluation? Different methods and metrics might be developed and trialled through interviews, surveys, controlled experiments, diary studies, and human annotation exercises.





A final set of theoretical considerations is concerned with the role of critical thinking as participating in a wider variety of end-user concerns. Research on the user experience of generative AI has highlighted that besides critical thinking, users also require support with checking the correctness of output \citep{gordon2023co}, with metacognitive activities (such as developing a mental model of their own capabilities in prompting generative AI) \citep{tankelevitch2023metacognitive}, and explanations for AI-generated output. It is highly likely that designing for each of these separately and then cobbling them together into a Frankenstein interface is a recipe for information overload. How can these varying aims and support needs be unified? For example, the grounded abstraction matching approach \citep{liu2023wants} shows how a single interface might serve ``double duty'' as an explanation mechanism, metacognitive support, and tutorial. Through design probes and experiments, we might begin to explore the key design trade-offs between critical thinking and other concerns, such as metacognition, explanation, task efficiency and correctness, information detail and frequency, and user fatigue.





\subsubsection*{Interface Design}

A second set of questions seeks to explore the design space of critical thinking interventions (notwithstanding observations that the ``design space'' framing can be counterproductive for HCI research by inappropriately applying scientistic principles and forms of rigour to the design discipline \citep{reeves2015human}).

Multiple types of content could be the target of critiques. For instance, as exemplified by the shortlisting prototype, we could focus our critiques on AI-generated content, on the principle that this content is the most likely candidate for participants to passively offload cognitive tasks onto, and thus the greatest threat to critical thinking. But this might be unnecessarily restrictive. We could critique content (data, formulas, text, etc.) that the user has generated themselves. We might even critique user queries to the system, on the principle that this represents an intent which might itself benefit from critical thinking.

There is the question of the kinds of information that can be presented in critiques. For example, a critique could focus on limitations of the content, or highlighting biases in the content, or risks in the content, or proposing alternatives to the content. Critiques may concern each level of the Bloom taxonomy such as analysing (organisation of ideas into patterns), or synthesising (composing, modifying, and predicting ideas), or evaluating (comparing and judging ideas). The user has a finite attention budget, and critiques can be (potentially) infinitely detailed, so what type of content should we prioritise? The quantity of information is also key: as previous research has shown, too much detail in AI explanations can overwhelm users \citep{kulesza2015principles,sarkar2022explainable,sarkar2024large} and a similar effect is likely to be observed in AI critiques as well.

Moreover, how can this information be best presented? Language models predispose us to think in terms of textual critiques, and these do provide an attractive and straightforward option, but critiques might be assembled and presented in a variety of visual and diagrammatic conventions, such as argument maps, or even as videos narrated by anthropomorphised characters.

Related to the concerns of information overload, there are questions of the dynamics of critical thinking interventions. How often should critiques be presented to users? Should they be persistent and ambient information, or proactively surfaced at certain intervals or during certain events? If so, what are these events, and are interventions equally effective if presented before, during, or after querying the system? Or should they be surfaced only on user request? Does the location onscreen of critiques have an impact -- e.g., can the choice of placing a critique adjacent to the content, versus within a conversation log with the AI agent, versus in a dedicated space for critiques, have a meaningful effect on outcomes?

What personalisation needs do critiques have? Is it sufficient to have an opinionated design that determines the target, content, form, timing, and location of critiques? If each of these were made configurable, does the added benefit of personalisation outweigh the added complexity of end-user configuration?

The design space of critiques is thus incredibly complex and has a high dimensionality, and it is unlikely that the dimensions outlined above are independent of each other. Moreover, any design must necessarily answer to \emph{every} question; as an ``ultimate particular'' \citep{stolterman2008nature}, it will be subject to innumerable design decisions and constraints, not all of which we can simultaneously have independent empirical rationale for. This creates a bootstrapping problem where in order to investigate any variable it would appear that we would first need to have understood all the others. This makes it challenging to evaluate each variable in isolation through controlled experiments. Instead, a plausible way forward is to develop an annotated portfolio \citep{gaver2012annotated} of different points in the design space, starting with lightly principled but not inviolable assumptions, grounded in particular user tasks, as a way of exploring tradeoffs and creating a basis for fine-grained controlled experimentation. These could be evaluated as design probes, but could also be suitable targets for large-scale surveys, if a sufficiently effective survey method for evaluating critical thinking outcomes is available.




















\subsection{Technical Implementation Questions}

Generation of critiques poses several technical challenges. How do we generate different types of critical thinking interventions, with different modes (e.g., text/image/diagram) at different levels of detail, with different forms (e.g. as questions), with different sources and strategies (e.g., using retrieval-augmented generation to adapt critiques to personal task or organisational data)? Answering these questions will involve detailed experimentation with prompt engineering and architectures.

Importantly, how do we ensure that these critiques are of high quality? How can we rapidly, perhaps automatically, evaluate the quality of generated critiques? Are there any ``objective'' metrics that can be used to expedite and improve the quality of evaluation, whether done automatically or through human annotation?



A final set of questions concerns the system architecture and cost and performance optimisations. How can we scale critical thinking interventions to provide increased responsiveness to the user, or reduce the cost of repeatedly invoking large language model inference? In particular the strategy of using multiple calls to language models to simulate multiple agents before providing an answer might improve quality but at a greatly increased cost in generation time. Already we are seeing design patterns emerge to help users navigate this trade-off, such as Bing's ``deep search'' feature,\footnote{\url{https://blogs.bing.com/search-quality-insights/december-2023/Introducing-Deep-Search}. Last accessed: 23 April 2024.} which is introduced with the following explanation: \emph{``Going deeper takes time [...] but it can be worth the wait for more specific or comprehensive answers.''} Interfaces that explicitly surface the trade-off between computational time and result accuracy have been previously explored \citep{sarkar2015uncertainty}.



\section{Conclusion}

This paper has explored the critical risk that generative AI poses to knowledge work -- that powerful AI capabilities may lead humans to disengage from critical thinking and simply defer to AI output. This risk of moving to ``autopilot'' is an even greater challenge than the more commonly discussed issue of AI hallucinations or factual errors. The more pernicious outcome is that generative AI becomes complicit in intellectual deskilling and the atrophy of human critical thinking faculties. 


To help foster critical thinking dispositions when using generative AI systems, we discuss the concept of developing AI experiences not just as capable assistants, but as provocateurs \citep{sarkar2024challenge} that intentionally expose users to contrasting viewpoints, surface risks and limitations, and stimulate consideration of alternative perspectives. We instantiated these ideas in a prototype system for the activity of spreadsheet-based shortlisting. Our prototype uses an AI model to generate suggested criteria for shortlisting data rows, but also generates textual ``provocations'' that critique and highlight potential biases or oversights in those criteria. While prior work has explored AI critiquing capabilities and argumentation tools to support reasoning, this prototype is one of the first to deeply integrate critiquing directly into a generative AI experience flow with the explicit goal of actively cultivating critical thinking as a design principle.

Implementing an AI provocateur is a novel interaction model that opens up a rich design space to explore. Key questions include when and how provocations should be delivered, what form they should take (e.g. text, visualizations, interactive widgets), whether they should be generated dynamically or draw from a curated repository, and how to balance provocation with a more assistive role. Key next steps include: creating a better understanding of the role of critical thinking in contemporary knowledge work; developing richer provocation generation techniques that go beyond textual descriptions to include visualizations, interactions, and multimedia; conducting studies to assess which provocation modalities, timing, and interaction flows are most effective for different users/tasks; and exploring adaptive approaches that personalize provocations based on analyzing the user's exploration process and responses.


Critical thinking is not just an isolated skill, but needs to become a core design principle for the next generation of AI user experiences across domains. Rather than automation or ambient intelligence, AI should facilitate what we call ``Tools for Thought'' \citep{iverson2007notation, sarkar2024challenge} -- tools that amplify and extend our analytical reasoning capabilities in a manner that maintains human agency. Realizing this vision will require combining insights spanning theories of education, models of reasoning and argumentation, generative AI techniques, human-centered design, and a deep commitment to shaping technology to cultivate critical thinking as an essential skill for the 21\textsuperscript{st} century.



\bibliographystyle{plainnat}
\bibliography{references}  

\clearpage

\appendix
\section{Implementation of the Critical Shortlisting Prototype}
\label{apx:implementation}
\subsection{Architecture}

The prototype is structured as a contemporary web app, and is divided into a React\footnote{\url{https://react.dev/reference/react}. Last accessed: 23 April 2024.} front end and a Python back end. To generate suggestions, provocations and analyses, the front end makes requests to the back end using REST API calls. The back end in turn constructs prompts to send to an LLM, and returns the LLM's responses to the front end. The prototype can be run locally or with remote control, and could in principle be deployed to run a diary study. We use LangChain\footnote{\url{https://python.langchain.com/docs/get_started/introduction/}. Last accessed: 23 April 2024.} to simplify prompting the LLM to operate on a dataset.






Factors are generated with a single API call to the back end. The back end constructs a prompt from the shortlist criteria and the first 40 dataset rows for context, and sends it to the LLM. The LLM responds with an array of factors in a well-defined schema, including the title, source columns, factor criteria, importance, and a provocation.


Factor analysis is likewise done using a single API call to the back end. The back end constructs a prompt from the factor criteria and the first 5 dataset rows, and sends it to the LLM. The LLM responds with a Python program to filter dataset rows, which the back end runs against the entire dataset.

The time taken for the LLM to generate responses dominates all performance considerations. Most take 15 to 60 seconds, which is long enough to be disruptive and frustrating to users. The delay is primarily due to back-and-forth chatter between LangChain agents and the LLM, and is outside our control. The implementation therefore adheres to the following principles.
\begin{itemize}
    \item Limit interactions with the LLM; e.g. don't have it do tasks that can be done with heuristics and fixed logic, such as generating statistics for numerical columns.
    \item Avoid making serial requests to the LLM; e.g. always generate provocations at the same time as suggestions.
    \item Use caching when possible in experiments.
\end{itemize}


\subsection{Automation and Caching}

Caching LLM-generated content is important for more than just performance. The amount of control that can be wielded over an LLM's output is varying and unpredictable. Model parameters meant to modulate output variation or creativity, such as temperature, cannot currently be used to force LLMs to behave deterministically. Further, LLMs available only as services can change their behaviour without warning. Thus, to improve repeatability and increase the speed of interaction with our prototype during the study, we introduced a caching mechanism whereby a dataset can be preloaded, and pre-prepared factor generations can be cached and retrieved without additional LLM calls. Cached LLM responses can also be modified for experimental purposes, for example, to manually adjust the brevity, tone, or quality of provocations to serve as different experimental variations or conditions.




Caching LLM-generated content introduces an opportunity to streamline experiments by automating user interactions in the prototype. For example, any cached factor generations are initially created with a specific dataset as context, so the prototype may as well load that dataset on the user's behalf.


\begin{figure}
\centering
\begin{small}
\begin{BVerbatim}
export interface Scenario {
  /** The scenario name shown in the drop-down and on the prototype canvas. */
  displayName: string;

  /** If defined, the prototype uses this dataset immediately. */
  autoUploadFilename?: string;

  /** If true, the prototype analyzes factors as soon as they're generated. */
  analyzeFactorsImmediately?: boolean;

  /** Implementation of factor generation. */
  criteriaToFactors: (criteria: string, ...) => Promise<Factor[]>;

  /** Implementation of factor analysis. */
  getAnalysisContent: (factorSource: ColumnNames, factorCriteria: string, ...)
    => Promise<GetAnalysisContentResponseBody>;

  /** Implementation of "Update Analysis". */
  getOverallRecommendation: (criteria: string, currentFactors: Factor[], ...)
    => Promise<GetOverallRecommendationResponseBody>;
}
\end{BVerbatim}
\end{small}
\caption{\texttt{Scenario} interface (in TypeScript). Implementations of \texttt{Scenario} are used to automate user interactions, and to intercept back end requests for caching and experimental control.}
\label{fig:scenario}
\end{figure}

For both automation and caching, we use dependency injection.\footnote{\url{https://en.wikipedia.org/wiki/Dependency_injection}. Last accessed: 23 April 2024.} The specific dependency to be injected is a \textit{scenario}, defined by the \texttt{Scenario} interface shown in Figure~\ref{fig:scenario}. On prototype startup, the experimenter selects a \texttt{Scenario} implementation by name, which thereafter determines the prototype's behaviour. In a user study, a well-defined study task which takes the form of choosing a shortlist of the top $n$ rows from a specified dataset with some general (albeit as-yet incomplete and underspecified) target criterion, each task can be cached as a particular \texttt{Scenario}.



An implementation of \texttt{Scenario} called \texttt{DefaultScenario} causes the prototype to behave with no automation or caching; i.e. the user does everything manually, and factor generation and analysis requests are sent to the back end. The default scenario also logs every back end response. To define a new scenario, a researcher can extend the default scenario by:
\begin{enumerate}
    \item Choosing the default scenario on prototype startup.
    \item Using the prototype as it is expected to be used in an experiment.
    \item Creating a new class that extends \texttt{DefaultScenario}, with methods that return the logged responses instead of making back end requests.
\end{enumerate}
The methods may return logged responses with or without alterations, and under whatever circumstances are relevant to the experiment. In principle, any user action can be automated and any back end request can be intercepted. It is not difficult to extend \texttt{Scenario} with more properties and methods, and to make the prototype's behaviour depend on them.


\subsection{Generating Provocations}


Provocations are generated along with the suggestions in the current prototype. Future implementations may use a separate LLM call to generate provocations after suggestions are generated.
For example, the factor generation prompt asks the LLM to return, as part of its JSON response, a \texttt{risk} property that is used directly as the provocation on the factor card. The relevant excerpt from the prompt:
\begin{center}
\begin{small}
\begin{BVerbatim}
Each factor must contain the following information:
- "name": The name of the factor or criteria.
- [...]
- "risk": The risk of using such criteria, and what alternative criteria could be used.
  Suggest more relevant topics and keywords to the factor description. Even if there
  would be no risk, suggest a case where the opposite of the criteria is better.
\end{BVerbatim}
\end{small}
\end{center}



Likewise, factor analysis provocations are generated concurrently. The relevant excerpt from the prompt:
\begin{center}
\begin{small}
\begin{BVerbatim}
Your answer must contain the following information:
- For each row, the row's "id_" and a "reason" to include the row.
- A "message" containing any warnings.
\end{BVerbatim}
\end{small}
\end{center}
The \texttt{reason} and \texttt{message} are displayed as-is on the factor analysis card.


In the current implementation, we instruct the LLM to generate provocations as part of the factor information. Further, all factors, including their descriptions and provocations, are generated together based on a single prompt. This can be improved using different prompting strategies which proved to be successful in similar scenarios. For example in explainable recommendations, recently different prompting strategies have been investigate to generate explanations~\citep{luo2024unlocking, Rahdari_2024} for setting similar to ours.  

While the prompting strategies for explanations show promising results, using LLMs to generate explanations have several limitations and risks which also apply to our scenario of generating provocations like biases~\citep{gallegos2024bias} or hallucinations~\citep{huang2023survey}. Further investigations need to be done to evaluate the quality of the generated provocations. While user-studies might show the quality of the provocations in end-to-end scenarios, they are not a viable option during the design and development phase due to time and costs. Approaches like LLM-as-a-judge~\citep{liu-etal-2023-g} can be a potential alternative for future investigation.

\end{document}